\documentclass{appolb}
\usepackage{graphicx}
\usepackage[multiple]{footmisc}
\usepackage{wrapfig}

\graphicspath{{./plots/}} 

\begin{document}
    \title{Recent quarkonium results at Belle II
    \thanks{Presented at Excited QCD, Giardini Naxos, Italy}}
    \author{
        {Dmytro Meleshko, Elisabetta Prencipe, Sören Lange
        \address{Justus-Liebig-Universität Gießen, Gießen, Germany}}
        \\[3mm]
        }
    \maketitle
    \begin{abstract}
        The Belle experiment has given a substantial contribution in the field of heavy quarkonia. Several new states were announced by Belle for the first time, in both charmonium and bottomonium spectrum, or the confirmation from Belle corroborated former observations. Belle II is a next-generation experiment, which aims to continue and expand further the Belle physics program. We report in this document about the current status of the experiment and early physics results. Particular attention is devoted to the results of the analyses of the data sets collected during the energy scan (Nov 2021). 
    \end{abstract}
      
    \section{Introduction}
    
    The B-factories Belle and BaBar shed a light on the Quarkonium physics. In fact, nowadays we count tens of new states, first observed at these $e^+e^-$ asymmetric colliders. 
    Not all observations showed good agreement with the predictions from theoretical quark models. Soon a plethora of new models raised, clearly showing the inadequacy of the former in classifying all observed states in a unique model.     
    B-factories like BaBar and Belle offer several advantages in searching for new states and rare decays, for example the setup of these electron-positron detectors allowed full-event reconstruction and offered a clean environment for this kind of physics. They always represented a complimentary tool for the search for exotic states, performed at proton-proton experiments.
      
    B-factories can profit from running at the nominal center-of-mass (CM) energy of the $\Upsilon(4S)$ resonance: $B \bar B$ pairs are produced at rest, generating enormous quantity of $b \bar b$, then $c \bar c$ pairs. This allows to perform quarkonium studies, both charmonium and bottomonium spectroscopy.

    In Belle II, which is a major upgrade of the Belle experiment, the energy of 10.58 GeV (and potentially it is allowed up to 11 GeV) can be reached as a unique physics feature of this experiment. On the other hand, the unique configuration of the Belle II detector gives a chance to study decay channels with soft (energy down to 30 MeV) neutral particles in the final state. This opens the possibility to study various production mechanisms, which, in combination with the large statistics to  be collected in the upcoming years, makes Belle II physics studies unique or complimentary compared to the results from current running experiments.
    
    One of the main features of the Belle II detector is the vertex detector (SVD) upgrade and the installation of the Pixel Detector (PXD). In particular, it  includes the installation of a 2-layer PXD and a 4-layer SVD detectors for better vertexing performance. The electromagnetic calorimeter (ECAL) upgrade is going, and was made to ensure better background tolerance and readout electronics. The High-Level-Trigger (HLT) upgrade deals with higher event rate ($\sim$ 30 kHz). Alongside with the detector upgrade, KEKB accelerator was upgraded to its SuperKEKB version. Its main feature is the so-called "nano-beams" technique, which allows a decreasing of the transverse dimension of the beam down to $nm$ scale. This technology in tandem with the factor 2 current increase will allow an upgrade in integrated luminosity, which is planned to be a factor 40 higher w.r.t. Belle. So far Belle II collected 430 fb$^{-1}$ data in roughly 3 years of data taking. This is the same size as the whole dataset of BaBar experiment and half of all Belle data, collected in less than half time.
    
    The Belle II experiment is living so far its first long shutdown, waiting for the major upgrade of the second PXD layer, which is supposed to happen in a few months. Before the shutdown, an upgraded accelerator and a partially upgraded detector happened. An important day for the collaboration is April 18, 2018, when the first collision took place, which started  the first data  collection phase. The target integrated luminosity of the Belle II experiment by the end of its operation is 50 $ab^{-1}$. 
    
    \section{Charmonium-like studies}
    
    At B-factories we can search for new $quarkonia$ in different production mechanisms. Without any doubt, the suppressed B decays are the most analysed for this purpose. The golden channel to be analysed in B decays is identified  in the $B\to K X_{c\bar{c}}$ processes. They appear to be CKM favored with relatively large branching fractions ($10^{-3} \sim 10^{-4}$). The absolute branching fractions of many of these decays are still unknown, though progresses have been done $e.g.$ for $B\to X(3872)K$. The same applies to the $charmonium-charged$ $Z$ states, still not confirmed by Belle II because of lack of statistics. We also state that $Z_c$ neutral partners were not observed yet, which makes the puzzles of $X_{c\bar{c}}$ and $Z_c$ worth revising, when more data is available.
    
    We devote particular attention to the most cited resonance discovered by Belle, the $X(3872)$. Its early search with limited Belle II data (62$fb^{-1}$) brought observation of $14.4\pm4.6$ events ($4.5\sigma$), which is considered to be the first $X(3872)$ observation in Belle II. At the same time, MC studies show 20\% $\psi(2S)$ reconstruction efficiency gain w.r.t Belle. But the most intriguing side of the new $X(3872)$ study at Belle II refers to its width measurement in $D^0\bar{D^0}\pi^0$ decays. The Belle experiment was the first to observe the $X(3872)$ in this decay mode. The advantage of the $D^0\bar{D^0}\pi^0$ mode compared to $J/\psi\pi^+\pi^-$ is a significantly smaller $Q$ value ($7.05\pm0.18$ vs. $495.65\pm0.17$ MeV). This allows to push the boundaries down, when it comes to mass resolution ($684\pm8$ keV vs. $1.93\pm0.04$ MeV), width UL measurement ($\approx280$ keV\footnote{with 50 $ab^{-1}$ of collected data and at 90\% CL}\footnote{the prediction of $X(3872)$ width provided by the Flat\'e fit is $220^{+70+11}_{-60-130}$ keV} vs. $0.96^{+0.19}_{-0.18}\pm0.21$ \cite{LHCb:3872width}) and systematic uncertainties (110 keV \cite{Belle:Dlifetime}).
    
    The study of the so-called $Y$ family decays into $c-$baryon pairs ($\Lambda^+_c,\Sigma^-_c$, $\Sigma^+_c\Sigma^-_c$) and $cs$-meson pairs ($D_sD_{s2}(2573)$, $D_sD^*_{s0}(2317))$ via initial state radiation (ISR) represents a unique physics case at Belle II. In particular, the well-known $Y(4260)$ state still needs to be rediscovered by Belle II. Our MC simulations, based on the Belle analyses,  predict 60 $Y(4260)$ events over an integrated luminosity of  100 fb$^{-1}$, which may open the room for new line-shape-study analysis. The Belle II program includes the analysis of other ISR processes, such as the search for $Z$ charged states in decays like  $Z_{cs}\to K^{\pm}J/\psi$, $D_s^-D^{*0}$+c.c., which were not observed at Belle.
    
    In addition, we underline that the $J^{PC}$ measurement of the $X(3915)$ decaying into $\omega J/\psi$ draws particular attention in two-photon processes. This is another way to study $quarkonia$ at B-factories, as well as the observation of $X(4350)$ decaying into $\phi J/\psi$. We also mention that  $e^+e^-\to(c\bar{c})_{J=1}(c\bar{c})_{J=0}$ production rule puzzle, and $J^{PC}$ of the $X(3940)$ measurement catches our eye on the double charmonium processes. 
    
    \section{Bottomonium-like studies}
    
    The CM energy of B-factories grants $Bottomonia$ a pole-position for studying various $\Upsilon$, $Y_b$ and $Z_b$ states. This was monetized in the observation of the $h_b(1P,2P)$ \cite{Belle:etab}, $\eta_b(2S)$ \cite{Belle:etaBANDhb}, $Z_b(10610,10650)^{\pm}$ \cite{Belle:ZbDiscovery}, and $Y(10753)$ \cite{Belle:Y10753} by Belle. 
    
    The Belle II bottomonium program also includes: the study of the $\Upsilon(5S)$ and the $\Upsilon(6S)$ discrepancy in $\pi\pi\Upsilon(nS)$ and $\eta\pi\pi$ decays, deeper study of the 10.750 GeV/c$^2$ energy region, and general $c\bar{c}$ and $b\bar{b}$ spectra discrepancy. 
    
\begin{figure}[htb]
        \vspace{-3mm}
        \centerline{%
        \includegraphics[width=0.35\textwidth]{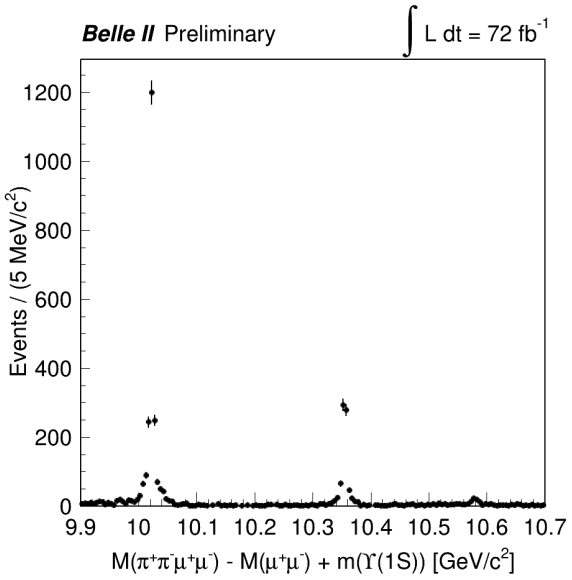}
        \includegraphics[width=0.35\textwidth]{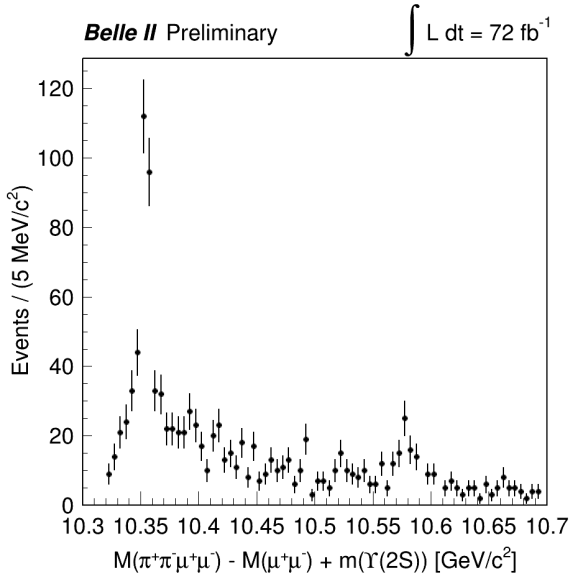}
        }
        \vspace{-1mm}
        \caption{Plots of $M(\pi^+\pi^-\mu^+\mu^-) - M(\mu^+\mu^-) + m(\Upsilon(nS))$ with a requirement of $|M(\mu\mu)-m(\Upsilon(nS))|<75$ MeV/c$^2$, where $m(\Upsilon(nS))$ represents the nominal $\Upsilon(nS)$ mass (n=1,2).}
        \label{fig:Fulsom_transitions}
        \vspace{-3mm}
    \end{figure}

    The first step to take in this direction has been the study of ISR events, and direct $\Upsilon(nS)\to\Upsilon(mS)$ transitions with early Belle II data (72 fb$^{-1}$). The result of this preliminary study brought to the confirmation of the transitions formerly seen in Belle, and to the first observation of the $\gamma_{ISR}\Upsilon(3S)\to\pi^+\pi^-\Upsilon(2S)$ transition with Belle II data, which was not seen in Belle at all (see. Fig. \ref{fig:Fulsom_transitions}). A precise $\Upsilon(4S)\to\pi^+\pi^-\Upsilon(nS)$ Dalitz analysis is ongoing to complement this picture. 
    
    Since the discovery of the $Y(10753)$ \cite{Belle:Y10753} in Belle energy scan data, its intrinsic nature remains still unclear. Its interpretation as a pure $\Upsilon(3S)$ state contradicts with theory \cite{Godfrey10753}, while its location on the mass scale makes unlikely its molecular interpretation. We do not even try to mention in this context its tetraquark and hybrid interpretations, for which a long debate has been ongoing. In addition to that, the similarity between $\pi\pi J/\psi$ and $\pi\pi\Upsilon$ cross-sections  \cite{Belle:Y10753, BESIII:JPsipipi} might indicate a common deflating mechanism. In order to raise the curiosity and the attention of the reader, we mention that the  $Y(4260)$ was observed by BESIII in $\gamma X(3872)$ and $\omega\chi_{c0}$ decays. So the logic question raises, if we should also expect $Y_b(10753)$ decaying into $\gamma X_b$, with $X_b\to\omega\Upsilon(1S)$.
    
    The Belle energy scan produced valuable and successful data analyses, so the same we expect in Belle II. In November 2021 a sequence of 5 runs at different CM energies in the range from 10.579 to 10.810 GeV was performed at SuperKEKB. During these runs, a total dataset of 15 fb$^{-1}$ has been accumulated (factor 4 larger w.r.t Belle). The 10.750 GeV/c$^2$ energy region has been revised with the new data. In the new study \cite{BelleII:10753}, two processes with the identical final-state products have been studied: $e^+e^-\to\omega\chi_{cJ}$ and $e^+e^-\to\gamma X_b$. In the first decay mode, 3 different $E_{CM}$ have been exploited ($\sqrt{s}=$ 10.701, 10.745, 10.805 GeV) in order to calculate cross-sections and complement them with the cross-section acquired by Belle (at $\sqrt{s}=$ 10.867 GeV). These 4 values are eventually enough to study $\sigma_b$ energy dependence. In the meantime, the second decay mode has been studied to search for the potential $X(3872)$ counterpart candidate, $e.g.$ $X_b$, at 4 different energies.

\begin{figure}[hbt]
        \vspace{-2mm}
        \centerline{\includegraphics[width=0.8\textwidth]{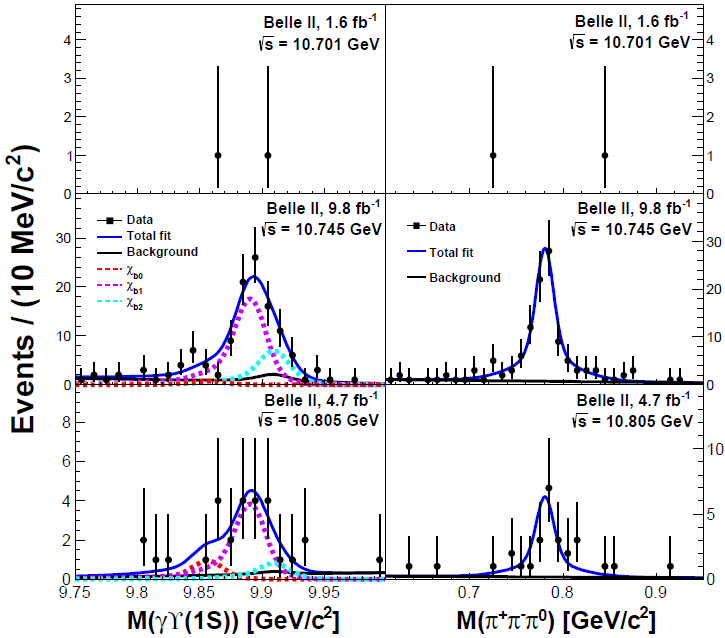}}
        \vspace{-1mm}
        \caption{Distribution of [left] $\gamma\Upsilon(1S)$ and [right] $\pi^+\pi^-\pi^0$ masses in data at $\sqrt{s}$=10,701, 10.745 and 10.805 GeV with fit results overlaid.}
        \label{fig:gammaUps_pipipi}
        \vspace{-5mm}
    \end{figure}

\begin{figure}[hbt]
        \vspace{-5mm}
        \centerline{%
        \includegraphics[width=0.7\textwidth]{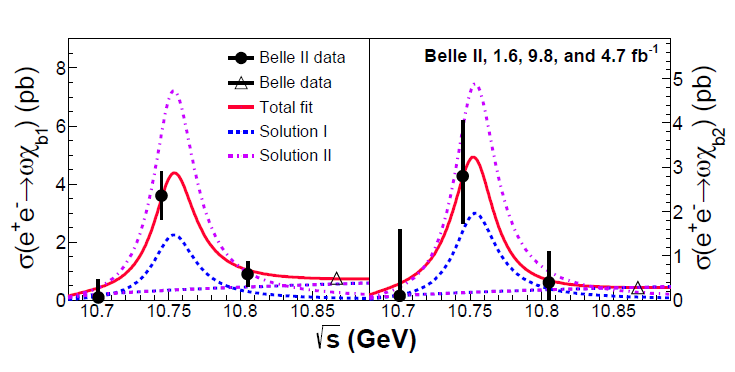}
        }
        \vspace{-3mm}
        \caption{Energy dependence of the Born cross-sections for $e^+e^-\to\omega\chi_{bJ}$. Circles show the Belle II measurements \cite{BelleII:10753}, triangles represent Belle results \cite{Belle:Y10753}.}
        \label{fig:cross-section}
        \vspace{-1mm}
    \end{figure}

    Study of the $e^+e^-$ $\to\omega\chi_{cJ}$ decay mode demonstrated significant signals (11$\sigma$ and 5$\sigma$) at $\sqrt{s}$ = 10.745 and 10.805 GeV (see Fig. \ref{fig:gammaUps_pipipi}). Based on the acquired cross-section values, one can calculate the values of $\mathcal{B}_f(Y_b(10753)\to\omega\chi_{b1})/\mathcal{B}_f(Y_b(10753)\to\pi^+\pi^-\Upsilon(1S))$ and $\mathcal{B}_f(\Upsilon(5S)\to\omega\chi_{b1})/\mathcal{B}_f(\Upsilon(5S)\to\pi^+\pi^-\Upsilon(1S))$ fractions, which appear to be one order of magnitude different. This may be considered as an unambiguous sign of the differences in the intrinsic nature of $Y_b(10753)$ and $\Upsilon(5S)$. The acquired cross-section values were complemented with the one from Belle and fitted with two solutions, the constructive and the destructive interference (see Fig. \ref{fig:cross-section}). The total fit function clearly indicates an enhancement near 10.753 GeV and do not hint to any structure in the vicinity of $\Upsilon(5S)$. Another result to mention here is the $\Gamma_{ee}\mathcal{B}_f(\Upsilon(10753)\to\omega\chi_{b1})/\Gamma_{ee}\mathcal{B}_f(\Upsilon(10753)\to\omega\chi_{b2})$ fraction, which appears to be close to 1. The result agrees with HQET predictions \cite{HQET}.

    Search for resonances in $\omega\Upsilon(1S)$ structure at different energies did not deliver $X_b$ observation, but only a distinctive $\omega\chi_{bJ}$ reflection is observed
    . We evaluate in any case $\sigma^{UL}_{X_b}$ upper limit, for each $E_{CM}$ and $M(X_b)$ in the energy range from 10.45 to 10.65 GeV.

    \section{Summary}
    
    B-factories started the XYZ studies, now complemented and pushed by studies covered from other experiments. Belle II is performing well, and will deliver new results that will shed the light on the most puzzling questions of heavy quarkonium. The long-term plan of the collaboration is to accumulate a data sample of 50 ab$^-1$. Early analyses at Belle II suggest possibilities for potential discoveries in the field of quarkonia.
    
    \bibliographystyle{unsrt}
    \bibliography{main}

\begin{thebibliography}{10}

\bibitem{LHCb:3872width}
Roel Aaij et~al.
\newblock {Study of the \ensuremath{\psi_2(3823)} and
  \ensuremath{\chi_{c1}(3872)} states in \ensuremath{B^+ \rightarrow \left(
  J\psi\pi^+\pi^-\right)K^+} decays}.
\newblock {\em JHEP}, 08:123, 2020.

\bibitem{Belle:Dlifetime}
F.~Abudin\'en et~al.
\newblock {Precise Measurement of the \ensuremath{D^0} and \ensuremath{D^+}
  lifetimes at Belle II}.
\newblock {\em Physical Review Letters}, 127(21), Nov 2021.

\bibitem{Belle:etab}
I.~Adachi et~al.
\newblock {First Observation of the P-Wave Spin-Singlet Bottomonium States
  $h_b$(1P) and $h_b$(2P)}.
\newblock {\em Physical Review Letters}, 108(032001), Jan 2012.

\bibitem{Belle:etaBANDhb}
R.~Mizuk et~al.
\newblock {Evidence for the \ensuremath{\eta_b(2S)} and Observation of
  \ensuremath{h_b}(1P)\ensuremath{\rightarrow}\ensuremath{\eta_b}(1S)\ensuremath{\gamma}
  and
  \ensuremath{h_b}(2P)\ensuremath{\rightarrow}\ensuremath{\eta_b}(1S)\ensuremath{\gamma}}.
\newblock {\em Phys. Rev. Lett.}, 109:232002, Dec 2012.

\bibitem{Belle:ZbDiscovery}
A.~Bondar, A.~Garmash, R.~Mizuk, D.~Santel, and K.~Kinoshita.
\newblock Observation of two charged bottomoniumlike resonances in
  $\ensuremath{\Upsilon}(5s)$ decays.
\newblock {\em Phys. Rev. Lett.}, 108:122001, Mar 2012.

\bibitem{Belle:Y10753}
R.~Mizuk et~al.
\newblock {Observation of a new structure near 10.75 GeV in the energy
  dependence of the $e^{+}e^{-}$\textrightarrow $\Upsilon(nS)$
  \ensuremath{\pi}$^{+}$\ensuremath{\pi}$^{-}$ (n=1,2,3) cross sections}.
\newblock {\em JHEP}, 10:220, 2019.

\bibitem{Godfrey10753}
Stephen Godfrey and Kenneth Moats.
\newblock Bottomonium mesons and strategies for their observation.
\newblock {\em Phys. Rev. D}, 92:054034, Sep 2015.

\bibitem{BESIII:JPsipipi}
M.~Ablikim et~al.
\newblock {Precise Measurement of the
  ${e}^{+}{e}^{\ensuremath{-}}\ensuremath{\rightarrow}{\ensuremath{\pi}}^{+}{\ensuremath{\pi}}^{\ensuremath{-}}J/\ensuremath{\psi}$
  cross-section at CM energies from 3.77 to 4.60 GeV}.
\newblock {\em Phys. Rev. Lett.}, 118:092001, Mar 2017.

\bibitem{BelleII:10753}
I.~Adachi et~al.
\newblock {Observation of $e^+e^-\to\omega\chi_{bJ}(1P)$ and search for $X_b
  \to \omega\Upsilon(1S)$ at $\sqrt{s}$ near 10.75 GeV}, 2022.

\bibitem{HQET}
A.~G. Grozin.
\newblock {Introduction to the Heavy Quark Effective Theory}, 1999.

\end{thebibliography}
\end{document}